\documentclass[prd,preprint,a4paper,superscriptaddress,nofootinbib,11pt]{revtex4}
\usepackage{amsmath,amsfonts,amssymb}
\usepackage{txfonts}
\usepackage{epsfig}
\usepackage{graphics}
\usepackage[T1]{fontenc}
\usepackage[safe]{textcomp}
\usepackage{dsfont}

\usepackage{csquotes}

\begin{document}
\begin{center}

{\bf  \Large Non-Commutative space-time and Hausdorff dimension}

\bigskip

\bigskip

Anjana V.{\footnote{e-mail: anjanaganga@gmail.com}} E. Harikumar {\footnote{e-mail: harisp@uohyd.ernet.in}} \\
School of Physics, University of Hyderabad, Central University P O, Hyderabad, Telangana, 500046, India \\ [3mm]

and A. K. Kapoor {\footnote{e-mail: akkapoor@iitbbs.ac.in}} \\
School of Basic Sciences, Indian Institute of Technology, Bhubaneswar, Odisha, 751007, India \\[3mm]

\end{center}
\setcounter{page}{1}
\bigskip

\begin{center}
 {\bf Abstract}
\end{center} 

We study the Hausdorff dimension of the path of a quantum particle in non-commutative space-time. We show that the Hausdorff dimension depends on the deformation parameter $a$ and the resolution $\Delta x$ for both non-relativistic and relativistic   quantum particle. For the non-relativistic case, it is seen that Hausdorff dimension is always less than two in the non-commutative space-time. For relativistic quantum particle, we find the Hausdorff dimension increases with the non-commutative parameter, in contrast to the commutative space-time. We show that non-commutative correction to Dirac equation brings in the spinorial nature of the relativistic wave function into play, unlike in the commutative space-time. By imposing self-similarity condition on the path of non-relativistic and relativistic quantum particle in non-commutative space-time, we derive the corresponding generalised uncertainty relation. 

\newpage
 
\section{Introduction}

One of the outstanding issues in physics is the question of gravity at high energies or equivalently at extremely short distance. There are various approaches to construct a quantum theory of gravity, such as string theory \cite{string}, asymptotically safe gravity \cite{ASG}, non-commutative geometry \cite{NC1, NC2, NC3}, loop gravity \cite{dim4}, causal dynamical triangulation \cite{dim5}. Many of these approaches indicate a change in space-time dimension at high energies \cite{dim1, dim2, dim3, dim4, dim5, dim6, dim7, dim8, dim9}. This change in the dimension, referred as dimensional flow has been analysed for various models. Some of these studies analysed the spectral dimension of $\kappa$-space-time, using return probability of diffusion process defined in the corresponding space-time \cite{AH1, AH2, AH3}. Some of these studies led to investigations of multi-scale space-time \cite{multi1, multi2}, fractional space-time and their relevance to quantum gravity. Another notion of dimension known as Hausdorff dimension which gives the dimensionality of paths travelled by quantum mechanical particles moving on such quantum space-time has been analysed in \cite{NN}. 

In this paper, we investigate the dimension of the path of a particle moving in space-time whose coordinates satisfy a Lie algebraic type commutation relation. The coordinates of this kappa-space-time obey the following commutation relations
\begin{equation}
 [\hat{x}_0, \hat{x}_i]=ia\hat{x}_i,~~~~~ [\hat{x}_i, \hat{x}_j]=0 .\label{nc}
\end{equation}
This space-time naturally appear in the context of loop gravity \cite{LQG} and deformed special relativity \cite{DSR}.

Dimensional flow in this $\kappa$-space-time has been studied by constructing and analysing diffusion equations in \cite{AH1, AH2, AH3}. The same has been analysed using different approaches in \cite{dim6, dim7, dim8, amelino}. In studying spectral dimension of $\kappa$-space-time, a modification of integration measure is used in \cite{dim8}, where as the same was calculated using the solution of $\kappa$-deformed diffusion equation \cite{AH1, AH2, AH3}. In both the approaches, it was noted that the spectral dimension reduces with the probe scale. 

In \cite{AW}, dimension of path of particle obeying Schrodinger equation was analysed. This is done by first defining the path of a non-relativistic quantum mechanical particle as follows; the expectation value of the position operator is measured at equal intervals of time $\Delta t$. Now joining the neighbouring points where the particle is localised at times $n \Delta t$ and $(n+1) \Delta t$ (where $n$ is an integer) and repeating this for all the intervals (i.e., for n = 1, 2, 3, ... ), one gets a curve made up of $(n-1)$ line segments. This curve is defined as the path followed by the quantum mechanical particle. This definition, naturally leads to dependence of the length of the path to the resolution of the measuring instrument. But one would like to have a definition of length of the path which is independent of the resolution of instrument. The path of quantum mechanical particle as defined above \cite{AW} is clearly an everywhere continuous but nowhere differentiable curve  which is known as Koch curve, in the study of fractals \cite{fractal1, fractal2}. Length of such fractal curves which is independent of the resolution of the measuring instrument has been introduced by Hausdorff and using this, dimension of quantum mechanical particle was shown to be 2 in \cite{AW}. For a curve whose length is $l$ as measured by an instrument of resolution $\Delta x$, Hausdorff introduced a new notion of length, $L_H$ as \cite{AW}
\begin{equation}
  L_H = l (\Delta x)^{D_H -1} \label{h_l}
\end{equation}
Here, $L_H$ is required to be independent of $\Delta x$. This fixes the parameter $D_H$ uniquely. This value of $D_H$ is called the Hausdorff dimension of the fractal curve or path of the quantum particle. This study has been generalised to the case of a particle governed by relativistic quantum mechanics in \cite{CF}. For the relativistic particle, the path was defined in \cite{CF}, in terms of Newton-Wigner operator \cite{NW}. It was shown that the Hausdorff dimension in the ultra-relativistic limit is 1, where as in the non-relativistic limit, this dimension is 2, as obtained by \cite{AW}. It was further shown in \cite{CF} that the length of the path as well as the Hausdorff dimension of the path is independent of the spinorial structure of the relativistic particle. In \cite{CF}, the Hausdorff dimension for a relativistic particle is shown to be 1. Thus we see that the Hausdorff dimension in non-relativistic regime (which is 2) is reduced to one by relativistic effects, in the commutative space-time.

In \cite{NN}, the approach of \cite{AW} has been adapted for calculating the Hausdorff dimension of a quantum space-time. The quantum nature of space-time was brought in by the introduction of a minimal length, which prevents localisation of particles below this length. The introduction of this minimal length changes the measure of integration in a non-commutative space-time. This change does modify the expectation value of the position operator calculated on such quantum space-time. It was shown in \cite{NN} that the Hausdorff dimension of the path is 2 as long as the resolution $\Delta x$ of the measurement is much above the intrinsic minimum length associated with the quantum space-time. But if the resolution is much below this minimal length, the Hausdorff dimension $D_H$ is $1-d$, where `$d$' is the topological dimension of the space-time. It was shown that the quantum nature of the space-time reduces the Hausdorff dimension from 2, it can be zero or even negative \cite{NN}.

In this paper, we study the Hausdorff dimension of the path of a particle moving in the $\kappa$-deformed space-time defined by eqn.(\ref{nc}). Note that the deformation parameter `$a$' has length dimension, and sets a minimum length scale below which one cannot probe or localise the particle. This length scale should also restrict the minimum value of the resolution of the measurement of position of the particle. These effects of deformation incorporated in our analysis is through the parameter `$a$' by using most general $\kappa$-deformed wave function in calculating the position expectation value of the particle moving in $\kappa$-space-time. We then generalise this method to study the Hausdorff dimension of path of a relativistic particle, 
governed by $\kappa$-deformed Dirac equation \cite{NCdirac}. Applying the self-similarity criterion on the path of non-relativistic and relativistic particle on non-commutative space-time, we then derive generalised uncertainty relation valid up to first order in the deformation parameter.
 
In the non-relativistic case, we find that the Hausdorff dimension of the quantum path decreases with increase in the deformation parameter. In the relativistic case, we find that the spinor nature of the wave function plays a crucial role, unlike in the commutative space-time. Also, in this case,  we find that the Hausdorff dimension increases with $a$ in the ultrarelativistic limit. Thus, contrary to commutative case, relativistic effects leads to an increase in $D_H$ in the non-commutative space-time. In the non-relativistic limit of quantum regime $D_H$ is discontinuous at $a = 3.639\times10^{-13} m$. In this case, $D_H$ first increases with increase in the deformation parameter $a$ and diverges as $a \rightarrow 3.639\times10^{-13} m$. We also note that above $a = 3.639\times10^{-13} m$, $D_H$ is negative and as $a$ decreases from higher value, $D_H$ decreases and diverges to $-\infty$ at $3.639\times10^{-13} m$ (see Fig.\ref{fig2}). This result is entirely in contrast with what we obtained in non-relativistic case. This difference is due to the spinorial nature of the wave function of relativistic particle. 

This paper is organized as follows. In the second section, the path of a non-relativistic particle moving in kappa-deformed space-time is analysed. Path of the quantum particle is constructed by measuring the position expectation values at different times separated 
by an interval $\Delta t$. We calculate the Hausdorff dimension by measuring the distance travelled by the particle and find that it depends on the deformation parameter. In section 3, we investigate the Hausdorff dimension of relativistic quantum path on $\kappa$-sapce-time and show that it depends on the spinor nature of the wave function, unlike in the commutative space. In section 4, we derive the generalised uncertainty relation valid for $\kappa$-space-time, by imposing self-similarity condition on the path of the quantum particle. The analysis of results and summary are presented in the last section.

\section{Dimension of non-relativistic quantum paths}

In this section, we study the Hausdorff dimension of the path of a non-relativistic quantum particle moving in kappa-deformed space-time. 
First we calculate the average length travelled by a quantum particle in time $T$ and using we obtain Hausdorff length
which is used in calculatng the Hausdorff dimension. In doing so, we have treated cases where particle has (i) zero average momentum, and 
(ii) non-zero average momentum, separately.

Consider a non-relativistic particle moving in the $\kappa$-spacetime. Its path is constructed by measuring the position expectation value of the particle, with resolution $\Delta x$, at different times $t_0$, $t_1 = t_0+\Delta t$, ..., $t_N = t_0+N\Delta t$, and joining these points by straight lines. The average distance travelled by the particle in time $\Delta t$ is $\langle \Delta l\rangle $, and the total distance travelled in time $T=N \Delta t$ is
\begin{equation}
  \langle l \rangle = N \langle \Delta l \rangle , 
\end{equation}
where N is the number of intervals. It should be noted that this measurement of length will depend on the spatial resolution $\Delta x$ of the measuring instrument and also on the deformation of spacetime characterised by the deformation parameter `$a$'.

The wave function of a quantum particle in $\kappa$-spacetime is of the form
\begin{equation}
  \Psi_{\Delta x, a} (\textbf{x}, \Delta t) = \frac{(\Delta x + a)^{\frac{3}{2}}}{\hbar ^3} \int_{\mathbb{R}^3}\frac{d^3 p}{(2 \pi)^{\frac{3}{2}}} f\left(\frac{\lvert \textbf{p} \lvert (\Delta x + a)}{\hbar} \right) e^{\frac{i}{\hbar} \textbf{p}.\textbf{x}} e^{- i\frac{\lvert \textbf{p}\lvert^2 \Delta t}{2\tilde{m} \hbar} }, \label{eqnpsi}
\end{equation}
where $\tilde{m}$ is the modified mass of the particle due to $\kappa$-deformation \cite{AKKEH, ZH, stjepan} and depends on the realization of $\kappa$-space-time coordinate. As shown below, Hausdorff dimension is independent of the mass in the non-relativistic case where as it depends on the mass in relativistic case. In writing this wave function, we have used the fact that particles position is localized within a region of size $\Delta x$ and in addition there is a modification due to the deformation of space-time which is incorporated through `$a$' dependence of the wave function. Note that in the commutative limit (i.e., $a \rightarrow 0$), we get back the wave function in the commutative space \cite{AW}. We next calculate the average distance travelled by the particle $\Delta l$ in time $\Delta t$ and using this find the total distance travelled in finite time $T = N \Delta t$. For this, we first restrict to path of particle with zero average momentum (i.e.,in the classical limit, the particle is at rest).

Using the redefinition $ \textbf{k}= \frac{ \textbf{p} (\Delta x + a)}{\hbar}$, the above wave function in eqn.(\ref{eqnpsi}) is re-written as 
\begin{equation}
  \Psi_{\Delta x, a} (\textbf{x}, \Delta t) = \frac{1}{(\Delta x + a)^{\frac{3}{2}}} \int_{\mathbb{R}^3}\frac{d^3 k}{(2 \pi)^{\frac{3}{2}}} f(\lvert \textbf{k} \lvert) e^{i \textbf{k}.\frac{\textbf{x}}{(\Delta x+a)}} e^{- i\lvert \textbf{k}\lvert^2 \frac{\hbar \Delta t} {2 \tilde{m}  (\Delta x +a)^2 }}. \label{psi}
\end{equation}
Now we calculate average distance travelled by the particle in time $\Delta t$ as
\begin{equation}
  \langle \Delta l\rangle = \int_{\mathbb{R}^3} d^3 x \lvert \textbf{x}\lvert \lvert \Psi_{\Delta x, a} (\textbf{x}, \Delta t) \rvert ^2. \label{avgdis}
\end{equation}
Substituting (\ref{psi}) in (\ref{avgdis}) and letting $\textbf{y} = \frac{\textbf{x}}{\Delta x+ a}$, one obtains
\begin{equation}
  \langle \Delta l\rangle = (\Delta x + a)\int_{\mathbb{R}^3} d^3 y \lvert \textbf{y}\lvert  \left\lvert \int \frac{d^3 k}{(2 \pi)^{\frac{3}{2}}} f(\lvert \textbf{k} \lvert) e^{i \textbf{k}.\textbf{y}} e^{- i\lvert \textbf{k}\lvert^2 \frac{\hbar \Delta t } {2 \tilde{m}  (\Delta x +a)^2 }} \right\lvert ^2.
\end{equation}
The form of $f(\lvert \textbf{k} \lvert)$ is chosen such that the particle is localized in a region of size $\Delta x$ and it should be noted that the specific form of $f(\lvert \textbf{k} \lvert)$ is not important in our calculation, as it is in the commutative case\cite{AW}. In this paper, we choose $f(\lvert \textbf{k} \lvert) =A e^{-(1+\frac{a^2}{\Delta x^2})\lvert \textbf{k} \rvert^2}$, so that the integration can be done exactly. Upon integrating over $k$ in the above equation, we find
\begin{eqnarray}
   \langle \Delta l\rangle = (\Delta x + a)\int_{\mathbb{R}^3} d^3 y \lvert \textbf{y}\lvert \frac{A^2}{8} \frac{1}{(\alpha^2+\beta^2)^{3/2}} e^{-\frac{ \alpha \lvert \textbf{y}\lvert^2}{2(\alpha^2+\beta^2)}} \label{a}
\end{eqnarray} 
where A is the normalisation constant of the function $f(\lvert \textbf{k} \lvert)$, $\alpha=(1+\frac{a^2}{\Delta x^2})$ and $\beta = \frac{\hbar \Delta t}{2 \tilde{m} (\Delta x+ a)^2}$. This choice of $f(\lvert \textbf{k} \lvert)$ incooperates the modification of the integration measure \cite{NN}, to take into account the feature of minimal length associated with quantum space-time. Re-expressing the above equation in polar coordinate, we obtain
\begin{eqnarray}
  \langle \Delta l\rangle &=& (\Delta x + a) \frac{A^2}{8} \frac{1}{(\alpha^2+\beta^2)^{3/2}}\int_{0}^{\infty} dr 4\pi r^2 r e^{-\frac{r^2 \alpha}{2(\alpha^2+\beta^2)}} \nonumber \\
  &=& A^2 \pi (\Delta x+ a)\frac{\sqrt{\alpha^2+\beta^2}}{\alpha^2}. \label{NR_l}
\end{eqnarray}
Following eqn.(\ref{h_l}), we find the Hausdorff length for the quantum path as 
\begin{eqnarray}
  \langle L_H \rangle  \propto \Delta x^{D_H-1} \frac{T\hbar}{ \tilde{m}  (\Delta x + a)} \frac{\sqrt{1+\left(\frac{2 \tilde{m} (\Delta x + a)^2}{\hbar \Delta t}\right)^2\left[1+\frac{a^2}{\Delta x^2}\right]^2}}{[1+\frac{a^2}{\Delta x^2}]^2}.
\end{eqnarray}
When $(\Delta x + a) \ll \sqrt{\hbar \Delta t /2\tilde{m}}$ the above equation reduces to 
\begin{eqnarray}
  \langle L_H \rangle &\propto& \Delta x^{D_H-2} \frac{T \hbar}{ \tilde{m} (1 + \frac{a}{\Delta x})}\frac{1}{[1+\frac{a^2}{\Delta x^2}]^2}. \label{a1}
\end{eqnarray}
Unlike in the commutative space-time, here we define Hausdorff dimension as the value of the parameter $D_H$ where 
\begin{eqnarray}
\frac{\partial \langle L_H \rangle }{\partial \Delta x} &=& 0.
\end{eqnarray}
A similar approach was taken in the study of Hausdorff dimension of quantum space-time in \cite{NN}. The above condition gives
\begin{eqnarray}
  D_H = 2-\frac{a}{a+\Delta x} - \frac{4 a^2}{\Delta x^2+a^2}, \label{nrd_h}
\end{eqnarray}
which in the commutative limit (a $\rightarrow$  0) reduces to 2, as expected \cite{AW}. Note that Hausdorff dimension $D_H$ is independent of the mass of the particle $\tilde{m}$. For $a \ll \Delta x$, we obtain $D_H \simeq  2$, which coincides with the result in the commutative space. But for $\Delta x \ll a$, $D_H = 2 - (1+\frac{\Delta x}{a})^{-1} - 4(1+(\frac{\Delta x}{a})^2)^{-1} \simeq -3+\frac{\Delta x}{a}$. i.e., below the minimal length $a$, we obtain Hausdorff dimension of the path of the quantum particle as approximately -3. 

\begin{figure}[h!]
\caption{ Hausdorff dimension as a function of $a$.}\label{fig1}
 \includegraphics[height=2 in, width=2in]{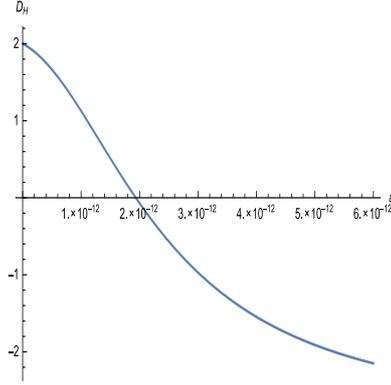}
\end{figure}

Taking $\Delta x$ to be the Compton wavelength of electron, i.e., $\Delta x = 2.426 \times 10^{-12} m $; we plot Hausdorff dimension as a function of $a$ in Fig.[\ref{fig1}]. From this figure, we see that in non-commutative space-time the Hausdorff dimension is always smaller than 2 and decreases with increasing deformation parameter $a$. The Hausdorff dimension attains negative value after a certain point. 

We know that in classical physics the Hausdorff dimension is always one and for a non-relativistic quatum particle it is greater than one \cite{AW, CF}. By requiring $D_H \geq 0$, we naturally obtain an upper bound on the deformation parameter which is of the order of $10^{-12}m$. For the case when $a\ll\Delta x$, we find that the Hausdorff dimension is less than 2 and for $a \gg \Delta x$ it is found that $D_H$ is -3. If the deformation is of the order of resolution (i.e., $a \simeq \Delta x$), we obtain $D_H$ as -0.5 (see enq.\ref{nrd_h}). Hausdorff dimension attaining zero or negative values, signals lose of meaning of space-time. This feature was also noted in \cite{NN} for a quantum space-time. Similar behaviour of spectral dimension was noted for non-commutative space-time and in particular $\kappa$-space-time \cite{AH1, AH2, AH3}.

Next we consider particle with non-zero average momentum (i.e., in the classical limit, the particle is moving) and calculate the Hausdorff dimension of its path. The wave function of a quantum particle with average momentum $\textbf{p}_{av}$ in deformed space-time is given by
\begin{equation}
  \Psi_{\Delta x, a} (\textbf{x}, \Delta t) = \frac{(\Delta x + a)^{\frac{3}{2}}}{\hbar ^3} \int_{\mathbb{R}^3}\frac{d^3 p}{(2 \pi)^{\frac{3}{2}}} f\left(\frac{\lvert \textbf{p} \lvert (\Delta x + a)}{\hbar} \right) e^{i \frac{(\textbf{p}+\textbf{p}_{av}).\textbf{x}}{\hbar}} e^{- i\frac{\lvert \textbf{p}+\textbf{p}_{av}\lvert^2 \Delta t}{2 \tilde{m}  \hbar}}.
\end{equation}
Proceeding as in the previous section with $\textbf{y} = \frac{\textbf{x}}{a+\Delta x}-\frac{\Delta t }{\tilde{m} (\Delta x +a)}\textbf{p}_{av} $ and $f(\lvert \textbf{k} \lvert) =A e^{-(1+\frac{a^2}{\Delta x^2})\lvert \textbf{k} \rvert^2}$, we find the length of the quantum path in $\Delta t$ time as
\begin{eqnarray}
  \langle \Delta l\rangle = (a+\Delta x)\int d^3 y \left \lvert \textbf{y}+\frac{\Delta t}{\tilde{m} (a+\Delta x)}\textbf{p}_{av}\right \lvert \left\lvert \int \frac{d^3 k}{(2\pi)^{\frac{3}{2}}} A e^{-\lvert \textbf{k}\lvert^2}e^{-i\lvert \textbf{k}\lvert^2 [\frac{\hbar\Delta t}{2\tilde{m}  (a+\Delta x)^2}-i\frac{a^2}{\Delta x^2}]} e^{i \textbf{k}.\textbf{y}}\right \lvert^2 \label{l_av}
\end{eqnarray}
Using the redefinition 
  \begin{equation}
 \frac{\hbar\Delta t}{2\tilde{m}  (a+\Delta x)^2} - i \frac{a^2}{\Delta x^2} = b, \label{b}
\end{equation}
where $b$ is demanded to be a constant, eqn.(\ref{l_av}) takes the form
\begin{eqnarray}
  \langle \Delta l\rangle = \frac{\Delta t \lvert \textbf{p}_{av}\lvert}{\tilde{m} }\int d^3 y \left \lvert \frac{\textbf{p}_{av}}{\lvert \textbf{p}_{av}\lvert}+\frac{ \hbar \textbf{y}}{2\lvert \textbf{p}_{av}\lvert(a+\Delta x)[b+i\frac{a^2}{\Delta x^2}]}\right\lvert \left\lvert \int \frac{d^3 k}{(2\pi)^{\frac{3}{2}}}A e^{-\lvert \textbf{k}\lvert^2}e^{-ib\lvert \textbf{k}\lvert^2} e^{i \textbf{k}.\textbf{y}}\right\lvert^2.
\end{eqnarray}
From this, using the definition of Hausdorff length ($\langle L_H\rangle $) in eqn.(\ref{h_l}), we find
\begin{eqnarray}
  \langle L_H \rangle = (\Delta x)^{D_H-1} \frac{T \lvert \textbf{p}_{av}\lvert}{\tilde{m} }\int d^3 y \left \lvert \frac{\textbf{p}_{av}}{\lvert \textbf{p}_{av}\lvert}+\frac{ \hbar \textbf{y}}{2\lvert \textbf{p}_{av}\lvert(a+\Delta x)[b+i\frac{a^2}{\Delta x^2}]}\right\lvert \left\lvert \int \frac{d^3 k}{(2\pi)^{\frac{3}{2}}}A e^{-\lvert \textbf{k}\lvert^2}e^{-ib\lvert \textbf{k}\lvert^2} e^{i \textbf{k}.\textbf{y}}\right\lvert^2. \label{hlav}
\end{eqnarray}
For the case, $\Delta x \gg \frac{\hbar}{\lvert \textbf{p}_{av}\lvert}$ (i.e., the resolution is much greater than the particle's wavelength), the Hausdorff length $\langle L_H \rangle$ is proportional to $(\Delta x)^{D_H-1}$ which sets  Hausdorff dimension to be 1, as in the classical case. In this limit, $D_H$ is independent of $a$.

When $\Delta x \ll \frac{\hbar}{\lvert \textbf{p}_{av}\lvert}$, we can neglect the term $\frac{\textbf{p}_{av}}{\lvert \textbf{p}_{av}\lvert}$ in comparison with the other terms in eqn.(\ref{hlav}). With $b=b_1+ib_2$, we obtain
\begin{equation*}
  \langle L_H \rangle \propto \frac{(\Delta x)^{D_H-2}}{(1+\frac{a}{\Delta x})}\frac{1}{\sqrt{b_1^2+(b_2+\frac{a^2}{\Delta x^2})^2}}
\end{equation*}
which gives the Hausdorff dimension as
\begin{eqnarray}
  D_H = 2-\frac{a}{a+\Delta x}-\frac{2\frac{a^2}{\Delta x^2}(b_2+\frac{a^2}{\Delta x^2})}{[b_1^2+(b_2+\frac{a^2}{\Delta x^2})^2]}. \label{nonzerodh}
\end{eqnarray}

Note that, in the commutative limit we get $D_H = 2$. From the above discussion, for path of a non-relativistic quantum particle, one can see that Hausdorff dimension is not an integer, and also it depends on deformation parameter $a$ as well as on resolution $\Delta x$. There are two `$a$' dependent correction terms in eqn.(\ref{nonzerodh}). For $a \ll \Delta x$ we find $D_H = 2$, which coincides with the commutative result. For $a\gg \Delta x$, we obtain $D_H = -1$. We also note that, $D_H$ for a particle with non-vanishing average momentum depends on the mass of the particle through b (see eqn.(\ref{b})).

\section{Dimension of relativistic quantum paths}

In this section, we present the calculation of Hausdorff dimension for relativistic Dirac particle in 
$\kappa$-space-time. Here we start with $\kappa$-deformed Dirac hamiltonian and re-express it in Foldy-Wouthuysen representation. This 
choice of representation help us to avoid non-physical observables. With this hamiltonian we solve the Ehrenfest equation to obtain the 
average length of a relativistic quantum particle traversing in $\kappa$-space-time during the time $\Delta t$. An interesting observation is 
that the distance travelled by 
the relativistic quantum particle in the $\kappa$-space-time depends on the spinorial character of the wave function,
in contrast to the commutative situation. The average 
length obtained is then used to calculate the Hausdorff dimension for various cases(such as classical limit, non-relativistic limit of quantum regime 
and Ultrarelativistic case).

The $\kappa$-deformed Dirac Hamiltonian, upto first order in $a$ is \cite{NCdirac}
\begin{eqnarray}
  H_{DNC} &=& H_D + \frac{ac}{2 \hbar} p^2,\label{H_DNC}
\end{eqnarray}
where $H_D = c \textbf{p}.\alpha +\beta m c^2 $. In the Dirac representation $\alpha$ and $\beta$ are gives as
\begin{equation}
\alpha_i = 
\left(\begin{array}{cc}
  0 & \sigma_i \\ \sigma_i & 0
\end{array}\right),
~~~~~~~\beta = 
\left(\begin{array}{cc}
  1 & 0 \\ 0 & -1
\end{array}\right)
\end{equation}
where $\sigma_i$ are the Pauli matrices.

Since we want to restrict our calculation to deal with only observables, we will be working in Foldy-Wouthuysen representation(FW-representation) \cite{Messiah}. The change from the Dirac representation to FW-representation is achieved by a unitary transformation implemented by
\begin{eqnarray}
   U = \frac{\beta H_D + E_p}{\sqrt{2E_p(mc^2+E_p)}}
\end{eqnarray}
where $E_p = (p^2 c^2 +m^2 c^4)^{1/2}$. The Hamiltonian in Foldy-Wouthuysen representation obtained by applying the above unitary transformation is
\begin{eqnarray}
  H_{FWNC} = U H_{DNC} U^\dag = E_p \beta +\frac{ac}{2 \hbar} p^2 , \label{H_phinc}
\end{eqnarray}
where $H_{DNC}$ is given in eqn.(\ref{H_DNC}). The modified position operator, $\textbf{X}$ compatible with the Foldy-Wouthuysen representation, known as Newton-Wigner operator \cite{Messiah} is 
 \begin{equation}
  \textbf{ X} =  U \textbf{x}U^\dag .\label{R} 
 \end{equation}
 Note that $\textbf{X}$  and $\textbf{p}$ satisfy the same commutation relation as $\textbf{x}$ and $\textbf{p}$, and thus we find $\Delta X \geq \hbar/\Delta p$. This $X$ satisfies the Ehrenfest equation 
 \begin{eqnarray}
  \frac{d^2 X^2}{dt^2} &=& -\hbar^{-2}[H_{DNC},[H_{DNC}, X^2]] \label{EE}
 \end{eqnarray}
which governes its time evolution.
 
Using (\ref{H_phinc}) and (\ref{R}) in (\ref{EE}), we obtain the equation for $X^2$ as
 \begin{equation}
 \frac{d^2 X^2}{dt^2} = \frac{2 c^4 p^2}{E_p^2} + \frac{4a}{\hbar} \frac{c^3 p^2}{E_p^2} H_D.
 \end{equation} 
By solving the above equation, we obtain the average value of $X^2$ at time t as
\begin{eqnarray}
  \langle X^2\rangle _t = c^4 \langle  \frac{p^2}{E_p^2}\rangle _0 t^2 + \frac{2ac^3}{\hbar} \langle \frac{p^2}{E_p^2} H_D\rangle_0 t^2 +\langle \frac{dX^2}{dt}\rangle _0 t + \langle X^2\rangle _0. 
\end{eqnarray}
where subscript $0$ on terms in RHS stands for values at t=0. The average length $\langle \Delta l \rangle$ travelled by the particle during the time interval $\Delta t \gg\langle \frac{dX^2}{dt}\rangle _0 /(c^4 \langle  \frac{p^2}{E_p^2}\rangle _0)$ will be  
 \begin{eqnarray}
   \langle \Delta l \rangle _{\Delta t} &=& ( \langle X^2 \rangle _{\Delta t} - \langle X^2 \rangle _0 )^{1/2} = c^2 \Delta t \langle \frac{p^2}{E_p^2}\rangle _0^{\frac{1}{2}} \left[ 1+ \frac{2 a}{\hbar c} \frac{ \langle \frac{p^2}{E_p^2} H_D\rangle_0}{\langle  \frac{p^2}{E_p^2}\rangle_0}\right]^{\frac{1}{2}} \label{rela len}
 \end{eqnarray}
Note that eqn.(\ref{rela len}) involves expectation value of $H_D$. Thus above expression for $\langle \Delta l \rangle$ shows that the distance travelled by the relativistic quantum particle depends on the spinorial character of the wave function. This should be contrasted with the commutative space-time result \cite{CF}, where the spinorial structure of the relativistic wave function does not play any role in the calculation of Hausdorff dimension. This is due to the fact that, in the commutative space $\langle \Delta l \rangle _{\Delta t}$ does not involve expectation value of $H_D$ and thus the result is independent of the spinorial nature of the wave function \cite{CF}. But here, we see that the spinorial nature of wave function comes into effect in the non-commutative space-time. Note that in the commutative limit, i.e., $a \rightarrow 0$, the $\langle \Delta l \rangle _{\Delta t} $ is independent of $H_D$ and thus all dependence of spinorial wave function drops out. In the momentum representation, the wave function of positive energy, spin up particle, compatible with non-commutative space-time (with minimal length given by $a$) is of the form
\begin{eqnarray}
  \Psi (p)_{NC} = \sqrt{\frac{E_p + mc^2}{2E_p}} \left(\begin{array}{c}
    1 \\ 0\\ \frac{c}{E_p +mc^2} p_z \\ \frac{c}{E_p + mc^2} (p_x +ip_y)
  \end{array} \right ) 
  \sqrt{\frac{3}{4 \pi}}\left(\frac{1}{\Delta p }+ \frac{a}{\hbar}\right)^{3/2} , \nonumber \\
 for \mid p - p_0 \mid <  \frac{\hbar \Delta p}{(\hbar+ a\Delta p)}
\end{eqnarray}
 
Using this wave function we obtain the average length $\langle \Delta l \rangle$ travelled by the particle as 

\begin{eqnarray}
  \langle \Delta l \rangle _{\Delta t} &=& c^2 \Delta t \left[ \langle \frac{p^2}{E_p^2}\rangle _0 +\frac{2a}{\hbar c}\langle \frac{p^2}{E_p^2} H_D\rangle_0 \right]^{\frac{1}{2}} \nonumber \\
  &=& c \Delta t \{1-\frac{3}{4} \left(1+\frac{a \Delta p}{\hbar}\right)^3 x^2 (\frac{2 \hbar}{a\Delta p+\hbar}-2x[\arctan\left(\frac{\frac{\hbar}{(a\Delta p +\hbar)}y + 1}{xy}\right)\nonumber \\ &+&\arctan\left(\frac{\frac{\hbar}{(a\Delta p +\hbar)}y - 1}{xy}\right)]+\frac{1}{2}\left[y\left((\frac{\hbar}{(a\Delta p+\hbar)})^2+x^2\right)-\frac{1}{y}\right] \nonumber \\ &\times& \log\left( \frac{(1+\frac{\hbar y}{a\Delta p +\hbar})^2+x^2 y^2}{(1-\frac{\hbar y}{a\Delta p +\hbar})^2+x^2 y^2}\right) ) + \frac{a}{\hbar} \frac{1}{(\Delta p)^3 p_0}(\frac{-9}{8}p_0 m^4 c^4[\log(-p_0 +\sqrt{p_0 ^2 +m^2 c^2}) \nonumber \\ &+& \log(p_0 +\sqrt{p_0 ^2 +m^2 c^2})] +\frac{1}{40} [\sqrt{(p_0 - \Delta p)^2 + m^2 c^2}(2p_0^4 - 11 p_0^2 m^2 c^2 +32 m^4 c^4 \nonumber \\ &+& 2 p_0^3 \Delta p -13p_0 m^2 c^2 \Delta p - 18 p_0^2 \Delta p^2 + 24 m^2 c^2 \Delta p^2 + 22 \Delta p^3 p_0 - 8 \Delta p^4) \nonumber \\&+& \sqrt{(p_0 + \Delta p)^2 + m^2 c^2}(-2p_0^4 + 11 p_0^2 m^2 c^2 - 32 m^4 c^4 + 2 p_0^3 \Delta p -13p_0 m^2 c^2 \Delta p \nonumber \\ &+& 18 p_0^2 \Delta p^2 - 24 m^2 c^2 \Delta p^2 + 22 \Delta p^3 p_0 + 8 \Delta p^4) ] +\frac{9}{8} p_0 m^4 c^4 \nonumber \\ &\times &[\log(-(p_0 -\Delta p)+\sqrt{(p_0 -\Delta p)^2+m^2 c^2})+\log((p_0 +\Delta p)+\sqrt{(p_0 +\Delta p)^2+m^2 c^2})] )\}^{1/2} \nonumber \\ \label{lrela}
\end{eqnarray}
where $x=mc/\Delta p$ , $y=\Delta p/p_0$.
 
From eqn.(\ref{lrela}), in the commutative limit ($a \rightarrow 0$), we obtain  
\begin{eqnarray}
 \langle \Delta l\rangle_{\Delta t} &=& c \Delta t \Bigg\{ 1- \frac{3}{4} x^2 \Big(2- 2x \left[ \arctan \left(\frac{y-1}{xy} \right ) + \arctan \left( \frac{y+1}{xy}\right)  \right]  \nonumber \\ &+&\frac{1}{2}[y(1+x^2)- y^{-1}] \log \left( \frac{(y+1)^2+x^2 y^2}{(y-1)^2+x^2 y^2} \right)\Big)\Bigg\}^{\frac{1}{2}}
\end{eqnarray} 
which is same as result obtained in \cite{CF}.

For $\Delta p \ll p_0$, i.e., in the classical limit, eqn.(\ref{lrela}) reduces to
\begin{eqnarray}
\langle \Delta l\rangle _{\Delta t} ^{cl} = c \Delta t \left[ \frac{p_0^2}{(p_0^2+ m^2 c^2)}+ \frac{2a}{\hbar}\frac{p_0^2}{\sqrt{(p_0^2+ m^2 c^2)}}\right]^{\frac{1}{2}}.
\end{eqnarray}

This quantity is independent of $\Delta X$. So the Hausdorff dimension $D_H$ is one as in the commutative case \cite{CF}.

For the case when $p_0 \ll \Delta p$, we consider two regimes, i.e.,

\begin{itemize}
   \item $\Delta p \ll mc$ (non-relativistic limit of quantum regime). In this case, eqn.(\ref{lrela}) reduces to
   \begin{eqnarray}
     \langle \Delta l\rangle _{\Delta t}^{QNR} = c \Delta t \left[ \frac{3}{5}\left(\frac{\Delta p}{mc}\right)^2\left(\frac{\hbar}{a\Delta p+\hbar}\right)^2 - \frac{3 a}{8 \hbar}\left(\frac{6 m^3 c^3}{(\Delta p)^2}+2mc\right)\right]^{1/2}. \label{NRQR_l}
   \end{eqnarray}
From this expression, we find the Hausdorff dimension to be
   \begin{eqnarray}
     D_H = 1+\frac{\frac{3}{5} (\frac{\hbar}{mc})^2\frac{\Delta X}{(a+\Delta X)^3}+\frac{9}{4} a (\frac{mc}{\hbar})^3(\Delta X)^2}{\left[\frac{3}{5}(\frac{\hbar}{mc})^2\frac{1}{(a+\Delta X)^2}-\frac{9}{4} a (\frac{mc}{\hbar})^3 (\Delta X)^2 - \frac{3}{4}\frac{amc}{\hbar}\right]}.
   \end{eqnarray}
We note that Hausdorff dimension depends on the mass of the particle, unlike the non-relativistic case (with $\textbf{p}_{av} = 0$) or the commutative case. Setting $\Delta X = \lambda_c$ (Compton wavelength of electron $\frac{\hbar}{mc}$), we plot $D_H$ against $a$ in Fig.\ref{fig2}.

\begin{figure}[h!]
\caption{ Hausdorff dimension as a function of $a$.}\label{fig2}
 \includegraphics[height=2in, width=2in]{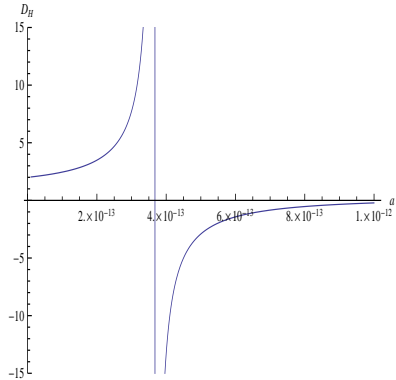}
\end{figure}

From the figure, we see that the Hausdorff dimension increases rapidly until $a$ approaches value of the order of $10^{-13}$ (for $\lambda_c = 2.426\times10^{-12} m$). At this point, the Hausdorff dimension blows up by approaching positive infinity and abruptly moves to negative infinity such that it becomes lesser negative for higher values of deformation parameter.

\item $\Delta p \gg mc$ (Ultrarelativistic case). In this case, eqn.(\ref{lrela}) become
  \begin{eqnarray}
    \langle \Delta l\rangle _{\Delta t}^{QR} = c \Delta t \left[ 1+ \frac{3}{2} \frac{a \Delta p}{\hbar} \right]^{1/2} \label{UQR_l}
  \end{eqnarray}
and the corresponding $D_H$ is  
  \begin{eqnarray}
   D_H = 1 + \frac{1}{2} \frac{3 a}{(2 \Delta X + 3 a)}.\label{ultradh}
  \end{eqnarray}
 \end{itemize}
 In the commutative space, the length travelled by the relativistic particle is independent of the resolution $\Delta X$. So the Hausdorff dimension is always one for the case of relativistic particle\cite{CF}. In our case, we see that the Hausdorff dimension depends on deformation parameter $a$ as well as resolution $\Delta X$, and it is always greater than one. This modification is due to the non-commutative nature of space-time. Also, note that $H_D$ is independent of the mass of the particle.
 
In the commutative limit, one can see that the distance travelled by the ultrarelativistic particle is independent of $\Delta p$ (or $\Delta X$). Hence Hausdorff dimension is always one in this limit.\\
\begin{figure}[h!]
\caption{ Hausdorff dimension as a function of $a$.}\label{fig3}
 \includegraphics[height=2in, width=2in]{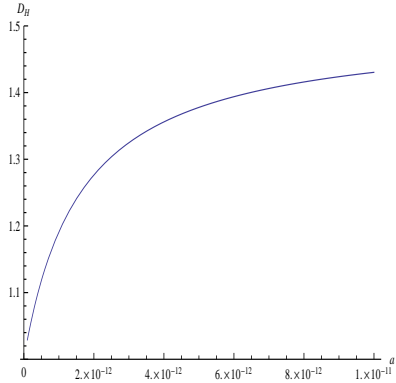}
\end{figure}
Hausdorff dimension given in eqn.(\ref{ultradh}) as a function of deformation parameter is plotted in Fig.\ref{fig3}. Here, we can see that the Hausdorff dimension increases with increase in deformation parameter and its value is always greater than one. 

In contrast to commutative case \cite{CF}, we see that the non-commutative nature of space-time results in an increase in the Hausdorff dimension. In the ultra-relativistic limit, Hausdorff dimension increases with $a$ while in the non-relativistic limit $D_H$ increases as $a$ increases and diverges to infinity. Beyond this critical value of $a$, Hausdorff dimension is negative as shown in Fig.\ref{fig2}.

\section{Modified Uncertainty Relation}

In this section we derive generalised uncertainty relations valid in the kappa-deformed space-time by using the concept of self-similar paths. We use energy-time uncertainty relation to find the generalised Heisenberg's uncertainty relation for non-relativistic and relativistic quantum particle.

In \cite{AW, CF}, it was shown that the self-similarity condition when applied to the path of the quantum particle, naturally leads to Heisenberg's uncertainty relation. Thus it is natural to anticipate that the requirement of self-similarity applied to path of a particle on non-commutative space-time would lead to modified uncertainty relation compatible with non-commutative space-time. The path of a quantum-mechanical particle will be self similar if $\langle \Delta l\rangle \propto \Delta x$. By demanding self-similarity for the path of a non-relativistic quantum particle (see eqn.(\ref{NR_l})), along with the condition $ (\Delta x + a) \ll \sqrt{\frac{\hbar \Delta t}{2 \tilde{m}}}$, we get the uncertainty in time as
\begin{eqnarray}
  \Delta t \simeq \frac{\tilde{m}}{\hbar} (\Delta x)^2 (1+\frac{a}{\Delta x}) (1+\frac{a^2}{\Delta x^2})^2.
\end{eqnarray}
Using this along with $\Delta E \sim \frac{(\Delta p)^2}{\tilde{m}}$ in $\Delta E ~\Delta t \geq \hbar$, we obtain the deformed uncertainty relation as 
\begin{eqnarray}
   \Delta x ~ \Delta p (1+\frac{a^2}{\Delta x^2}) \sqrt{1+\frac{a}{\Delta x}} \geq \hbar.
\end{eqnarray}
Keeping terms upto first order in $a$, we write the modified uncertainty relation in the form
\begin{eqnarray}
  \Delta x ~\Delta p (1+\frac{a}{2\Delta x}) \geq \hbar.\label{MUR_NR}
\end{eqnarray}
Following a similar procedure, using eqn.(\ref{NRQR_l}), for a spin half particle in the non-relativistic limit, we arrive at a modified uncertainty relation
\begin{eqnarray}
   \Delta X ~ \Delta p ~~(1.136)~~ \left[ 1+\frac{a}{2 \Delta X}+\frac{5}{16} a \Delta X^2 \left(\frac{mc}{\hbar}\right)^3\left(1+3\left(\frac{mc}{\hbar}\right)^2\Delta X^2\right)\right] \geq \hbar.\label{MUR_NRR}
\end{eqnarray} 
The corresponding uncertainty relation for a spin half particle in the relativistic regime ($\Delta E = c \Delta p $), derived from eqn.(\ref{UQR_l}) is
\begin{eqnarray}
 \Delta X~ \Delta p (1-\frac{3}{4}\frac{a}{\Delta X}) \geq \hbar.\label{MUR_R}
\end{eqnarray}
Note that eqn.(\ref{MUR_NR}) and eqn.(\ref{MUR_R}) show the generalised uncertainty relation, valid upto first order in $a$ is of the form $\Delta X ~\Delta p~ f\left(\frac{a}{\Delta X}\right) \geq \hbar$. The modified uncertainty relation obtained for non-relativistic limit of relativistic particle have other $a$ dependent terms apart from the $\frac{a}{\Delta X}$ term. Since the non-relativistic limit of relativistic path retains the information about the spinorial nature of the particle, it is natural that the result derived from this limit given in eqn.(\ref{NRQR_l}) is different from the result obtained by analysing the eqn.(\ref{NR_l}) for non-relativistic particle directly. Generalised uncertainty relations and their implications have been studied in \cite{GUP1, GUP2, GUP3}. Generalised uncertainty relation for a specific realisation of $\kappa$-deformed space-time has been discussed in \cite{GUP4}, where the modified uncertainty relation was found to have a mass dependence. We have shown that the self-similarity condition imposed on the path of a particle in non-commutative space-time leads to modified uncertainty relation. Also note that, the above obtained uncertainty relations gives the correct commutative limit.

\section{Conclusion}

In this paper, we have calculated the Hausdorff dimension for a quantum  particle moving in a non-commutative space-time. For a non-relativistic quantum particle, we found that the Hausdorff dimension is decreasing with deformation parameter $a$ for a fixed resolution $\Delta x$ (see Fig.\ref{fig1}). When the order of magnitude of deformation parameter is much smaller than the resolution of the apparatus (i.e., $a \ll \Delta x$), the Hausdorff dimension of the quantum path is practically two. This is also consistent with the result obtained in the commutative situation. In the contrasting case, where the resolution is of the same order or much smaller than $a$, we see that Hausdorff dimension approaches negative values, which do not yield a valid physical interpretation. Since, only a positive Hausdorff dimension provides the correct physical interpretation, demanding positivity of $D_H$ will provide us with an upper cutoff for the deformation parameter $a$ ($a < 10^{-12}m$ when $\Delta x \simeq$ compton wavelength of electron).  

An interesting result of this paper is that the path travelled by a relativistic quantum particle depends on the spinorial character of the wave function, unlike in the commutative space-time. It also depends on the mass of the particle for the relativistic case as well as for the non-relativistic case with $\textbf{p}_{av} \neq 0$. In contrast to the case of non-relativistic quantum path, Hausdorff dimension increases with the deformation parameter $a$ for a fixed $\Delta x$ and always takes positive values for a relativistic quantum path. When the order of magnitude of deformation parameter is much smaller than the resolution of the measurement, the Hausdorff dimension approaches one as in the commutative situation. We also note that the Hausdorff dimension of path of a relativistic particle is larger than one in the $\kappa$-deformed space-time which should be contrasted with the value of $D_H$ in the commutative space where it is always one. 

By imposing self-similarity condition on the path of non-relativistic and relativistic quantum particle in non-commutative space-time, we have derived the generalised uncertainty relation valid for $\kappa$-space-time. We found that due to kappa-deformation, the uncertainty relation is modified and the modification depends on a factor of the form $\frac{a}{\Delta x}$. All the obtained modified uncertainty relations, in the commutative limit, reduce to well known Heisenberg's uncertainty relation.

{\noindent{\bf Acknowledgements}}: AV thanks UGC, India, for support through BSR scheme. EH thanks SERB, Govt. of India, for support through EMR/2015/000622.

\end{document}